\documentclass[
11pt
              ,acmsmall
              ,nonacm
              ]{acmart}

\usepackage{amsmath,amsthm,amsbsy}
\usepackage{microtype}
\usepackage{algorithm}
\usepackage{setspace}
\usepackage{color}
\usepackage{xcolor}
\usepackage[noend]{algpseudocode}
\usepackage{multirow}
\usepackage{paralist}
\usepackage{xspace}
\usepackage[font=small,skip=0pt]{caption}
\usepackage{subcaption}

\makeatletter
\let\originalleft\left
\let\originalright\right
\renewcommand{\left}{\mathopen{}\mathclose\bgroup\originalleft}
\renewcommand{\right}{\aftergroup\egroup\originalright}
\makeatother

\newcommand{\set}[1]{\{ #1 \}}

\newcommand{\lb}{\left}
\newcommand{\rb}{\right}

\makeatletter
\newtheorem*{rep@theorem}{\rep@title}
\newcommand{\newreptheorem}[2]{%
\newenvironment{rep#1}[1]{%
\def\rep@title{#2 \ref{##1}}%
\begin{rep@theorem}}%
{\end{rep@theorem}}}
\makeatother
\newreptheorem{lemma}{Lemma}
\newreptheorem{theorem}{Theorem}

\newboolean{short}
\setboolean{short}{false}
\newcommand{\onlyShort}[1]{\ifthenelse{\boolean{short}}{#1}{}}
\newcommand{\onlyLong}[1]{\ifthenelse{\boolean{short}}{}{#1}}

\theoremstyle{definition}

\newtheorem{openproblem}{Open Problem}
\theoremstyle{plain}
\newtheorem{observation}{Observation}

\usepackage{mathtools}

\usepackage{tikz}
\usetikzlibrary{
  graphs,
  graphs.standard,
  fit,
  decorations.pathmorphing
}

\usepackage{algorithm}
\usepackage{algpseudocode}

\setboolean{short}{true}

\newcommand{\clique}{\ensuremath{\mathsf{UCC}}} %

\newcommand{\ktone}{\ensuremath{\mathsf{KT_1}}}

\newcommand{\DSketch}{\mathsf{SKETCH}}
\newcommand{\sketch}{\DSketch}

\usepackage[most]{tcolorbox}
\tcbuselibrary{theorems}
\newtcbtheorem[number within=section]{Definition}{Definition}{
  enhanced,
  sharp corners,
  attach boxed title to top left={
    yshifttext=-1mm
  },
  colback=white,
  colframe=gray!75!white,
  fonttitle=\bfseries,
  boxed title style={
    sharp corners,
    size=small,
    colback=gray!75!white,
    colframe=gray!75!white,
  }
}{def}
\tcolorboxenvironment{theorem}{lower separated=false,
                colback=white!90!gray,
colframe=white, fonttitle=\bfseries,
colbacktitle=white!50!gray,
coltitle=black,
enhanced,
boxed title style={colframe=black},
}
\tcolorboxenvironment{reptheorem}{lower separated=false,
                colback=white!90!gray,
colframe=white, fonttitle=\bfseries,
colbacktitle=white!50!gray,
coltitle=black,
enhanced,
boxed title style={colframe=black},
}

\title{What Can We Compute in a Single Round of the Congested Clique?}
\author{Peter Robinson}
\affiliation{
  \department{School of Computer \& Cyber Sciences}
  \institution{Augusta University}
  \country{Georgia, USA}
}

\ccsdesc[500]{Theory of computation~Distributed algorithms}
\keywords{Congested clique, distributed graph algorithm, lower bound}

\begin{document}
\begin{abstract}
We show that any one-round algorithm that computes a minimum spanning tree (MST) in the \emph{unicast} congested clique must use a link bandwidth of $\Omega(\log^3 n)$ bits in the worst case.
Consequently, computing an MST under the standard assumption of $O(\log n)$-size messages requires at least $2$ rounds.
This is the first round complexity lower bound in the unicast congested clique for a problem where the output size is small, i.e., $O(n\log n)$ bits. 
Our lower bound holds as long as every edge of the MST is output by an incident node. 
To the best of our knowledge, all prior lower bounds for the unicast congested clique either considered problems with large output sizes (e.g., triangle enumeration) or required \emph{every} node to learn the entire output.

\end{abstract}
\maketitle

\section{Introduction} \label{sec:intro}

In this work, we investigate the necessary link bandwidth for computing a Minimum Spanning Tree (MST) in the congested clique model~\cite{LPPP-spaa03}, where there are $n$ nodes that are associated with the vertices of an input graph $G$.
The nodes can communicate with each other over the links of the clique-topology in synchronous rounds, and the goal is to compute the solution to a graph problem on $G$.
We consider the powerful \emph{unicast congested clique}, where each node can send a (possibly distinct) message to every other node in the network.
Our work aims at understanding the computational power of the congested clique with respect to algorithms that compute their output based on the received messages after just a single round, while assuming that nodes have access to a shared source of randomness.  
We use $\clique_1$ to denote the one-round variant of the unicast congested clique.

We assume that the nodes are assigned unique IDs chosen from $\set{1,\dots,n}$.
In addition, we equip the nodes with a crucial amount of initial knowledge of the input graph, in the sense that each node knows not only its own ID but also the IDs of all of its neighbors, which is known as the \emph{$\ktone$ assumption} in previous literature (see \cite{DBLP:conf/podc/KingKT15,DBLP:journals/jacm/AwerbuchGPV90,DBLP:conf/podc/PaiPP021,DBLP:conf/soda/Robinson21}).

\section{Our Contributions and Related Work} \label{sec:contrib}

The problem of computing a minimum spanning tree is, without doubt, one of the most-studied graph problems in distributed computing and, in particular, in the congested clique model.
Here, we focus our discussion on the prior work in the congested clique and refer the reader to the survey of \cite{DBLP:journals/eatcs/Pandurangan0S18} for a more comprehensive discussion of distributed MST algorithms in general networks.
\cite{LPPP-spaa03} gave a deterministic $O(\log \log n)$-round algorithm, and, more recently, \cite{hegeman2015toward} used the linear graph sketches of \cite{AGM-soda12} to obtain a $O(\log\log\log n)$ -round randomized algorithm.
Subsequently, \cite{GP-podc16} showed that $O(\log^*n)$ rounds are possible by using sketches that are sensitive to the node degrees.
Shortly after, \cite{JN-soda18} presented an $O(1)$-round randomized algorithm, and \cite{DBLP:conf/stoc/000221a} showed that $O(1)$ rounds are  possible even deterministically.
These results crucially rely on the fast routing algorithm of \cite{L-podc13}, which itself requires $16$ rounds (see Theorem~3.7 in \cite{L-podc13}), and it is unclear whether any of these algorithms can be adapted to requiring just a single round, without increasing the necessary link bandwidth to $\Omega(n)$ bits.

\newcommand{\thmMST}{%
Any randomized algorithm that, with probability at least $\tfrac{3}{4}$, computes an MST on an $n$-node graph in a single round of the unicast congested clique, requires a link bandwidth of $\Omega(\log^3 n)$ bits.
This holds even if nodes have access to shared randomness, and when each edge of the MST is output by at least one of its two endpoints.
}
\begin{theorem} \label{thm:mst}
  \thmMST
\end{theorem}

An immediate consequence of Theorem~\ref{thm:mst} is that computing an MST requires at least two rounds under the standard assumption of limiting the link bandwidth to $O(\log n)$ bits.
We are not aware of any previous round complexity lower bounds for problems in the \emph{unicast} congested clique where each node may output only part of the solution and the total size of the output is $O(n\log n)$ bits, which is significantly smaller than the available per-round bandwidth of $O(n^2\log n)$.

The lack of existing lower bounds comes as no surprise: the seminal work of \cite{DKO-podc14} proved that lower bounds on the required number of rounds in the (multi-round) unicast congested clique is notoriously difficult for problems with small output sizes, as even slightly super-constant lower bounds (i.e.\ $\Omega(\log \log n)$) would yield new lower bounds on the circuit complexity of threshold gates.
Consequently, the only existing lower bounds for the unicast congested clique are shown for problems with large outputs.
For instance, in previous work~\cite{pandurangan2021distributed}, we showed a lower bound of $\tilde\Omega(n^{1/3}/\log^3 n)$ rounds for triangle listing, which, subsequently, was improved by Izumi and Le Gall~\cite{ig-podc17} to $\tilde\Omega(n^{1/3}\log n)$ rounds.
We point out that triangle listing has an output size of $\tilde\Omega(n^{7/3})$ in some graphs, and hence the round complexity bound stems from the fact that the per-round bandwidth is limited to $\tilde O(n^2)$ bits in the congested clique.
 
The work of \cite{MPRT-jcss20} gives exact bounds on the time complexity of the graph reconstruction problem in the \emph{broadcast} congested clique, under the assumption that {every} node must output all edges of the input graph. In particular, they show that reconstructing the edges of an arbitrary input graph $G$ requires $2$ rounds in the broadcast congested clique.
However, we note that their result does not hold under the relaxed assumption that each edge that is part of the solution must be output by \emph{some} incident node, i.e., the solution to the problem is the union of the nodes' outputs, which is what we assume in Theorem~\ref{thm:mst}.

\section{Proof of Theorem~\ref{thm:mst}} \label{sec:mst}

In this section, we prove that computing an MST in just one round of the unicast congested clique model is impossible under the standard assumption of $O(\log n)$ bits per message.

\definecolor{myblue}{RGB}{80,80,160}
\definecolor{vcol}{RGB}{196,218,238}
\definecolor{orange}{RGB}{225,113,57}
\definecolor{mygreen}{RGB}{211,238,205}

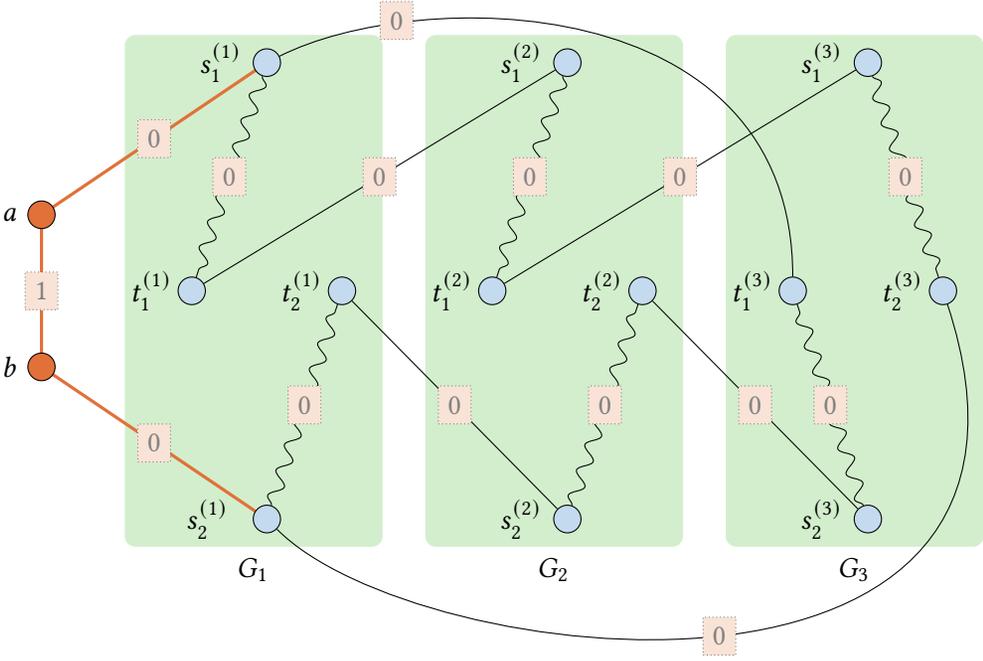
\begin{figure*}[t]
\begin{tikzpicture}[auto,yscale=1.00]
  \tikzset{vertex/.style={draw,circle,fill=vcol}} %
  \tikzset{vertex-set/.style={fill=mygreen,rounded corners}} %
  \tikzset{label/.style={text=black}} %
  \tikzset{long/.style={decorate,decoration=snake}} %
  \tikzset{edge/.style={draw,black}} %
  \tikzset{weight/.style={draw=gray,densely dotted,gray,fill=orange!20,midway,anchor=center}} %
  \tikzstyle{orange_edge}=[very thick,draw=orange,-,auto,in=0]
  \pgfdeclarelayer{background} \pgfdeclarelayer{inbetween} \pgfdeclarelayer{nodelayer} \pgfdeclarelayer{edgelayer}
  \pgfsetlayers{background,inbetween,edgelayer,nodelayer,main}
  \begin{pgfonlayer}{nodelayer} \node [style=vertex,fill=orange] (a) at (0, 7) {}; \node [left] at (a.west) {$a$}; \node [style=vertex,fill=orange] (b) at (0, 5) {}; \node [left] at (b.west) {$b$}; \node [style=vertex] (s11) at (3, 9) {}; \node [left] at (s11.west) {$s_1^{(1)}$}; \node [style=vertex] (s12) at (3, 3) {}; \node [left,xshift=-5.0] at (s12.west) {$s_2^{(1)}$}; \node [style=vertex] (t11) at (2, 6) {}; \node [left,xshift=2] at (t11.west) {$t_1^{(1)}$}; \node [style=vertex] (t12) at (4, 6) {}; \node [left,xshift=2] at (t12.west) {$t_2^{(1)}$}; \node [style=vertex] (s21) at (7, 9) {}; \node [left] at (s21.west) {$s_1^{(2)}$}; \node [style=vertex] (s22) at (7, 3) {}; \node [left] at (s22.west) {$s_2^{(2)}$}; \node [style=vertex] (t21) at (6, 6) {}; \node [left,xshift=2] at (t21.west) {$t_1^{(2)}$}; \node [style=vertex] (t22) at (8, 6) {}; \node [left,xshift=2] at (t22.west) {$t_2^{(2)}$}; \node [style=vertex] (sn1) at (11, 9) {}; \node [left] at (sn1.west) {$s_1^{(3)}$}; \node [style=vertex] (sn2) at (11, 3) {}; \node [left] at (sn2.west) {$s_2^{(3)}$}; \node [style=vertex] (tn1) at (10, 6) {}; \node [left,xshift=2] at (tn1.west) {$t_1^{(3)}$}; \node [style=vertex] (tn2) at (12, 6) {}; \node [left,xshift=2] at (tn2.west) {$t_2^{(3)}$}; \end{pgfonlayer}
  	\begin{pgfonlayer}{edgelayer} \draw[orange_edge] (a.center) -- node[left,weight,thin] {$1$} (b.center); \draw[orange_edge] (a.center) -- node[near start,weight,thin] {$0$} (s11.center); \draw[orange_edge] (b.center) -- node[weight,thin,midway,anchor=center] {$0$} (s12.center); \draw[long] (s11) to node[weight] {$0$} (t11); \draw[long] (s12) to node[weight] {$0$} (t12); \draw[long] (s21) to node[weight] {$0$} (t21); \draw[long] (s22) to node[weight] {$0$} (t22); \draw[long] (sn1) to node[weight] {$0$} (tn2); \draw[long] (sn2) to node[weight] {$0$} (tn1);

      \draw[edge] (t11) to node[weight] {$0$} (s21);
      \draw[edge] (t12) to node[weight] {$0$} (s22);
      \draw[edge] (t21) to node[weight] {$0$} (sn1);
      \draw[edge] (t22) to node[weight] {$0$} (sn2);
      \draw[edge] (tn1)
           .. controls (10,10) and (5,10) ..
           node[xshift=-8,weight,near end] {$0$} (s11);
      \draw[edge] (tn2)
           .. controls (14,0) and (5,1) ..
           node[above,weight] {$0$} (s12);
  	\end{pgfonlayer}

  \begin{pgfonlayer}{inbetween}
    \node[vertex-set,xshift=-5,inner xsep=15,inner ysep=5,fit=(s11) (s12) (t11) (t12)] (G1) {};
    \node[below] at (G1.south) {$G_1$};
    \node[vertex-set,xshift=-5,inner xsep=15,inner ysep=5,fit=(s21) (s22) (t21) (t22)] (G2) {};
    \node[below] at (G2.south) {$G_2$};
    \node[vertex-set,xshift=-5,inner xsep=15,inner ysep=5,fit=(sn1) (sn2) (tn1) (tn2)] (Gn) {};
    \node[below] at (Gn.south) {$G_3$};
  \end{pgfonlayer}

\end{tikzpicture}
\caption{%
\normalsize
The hard input graph $H$ on which computing an MST requires messages of $\Omega(\log^3 n)$ bits for one-round algorithms in the unicast congested clique.
The green-shaded regions are the blocks $G_1$, $G_2$, and $G_3$, which form the lower bound graph $G$ of \cite{Y-soda21}.
(Note that the figure shows a simplified version of this graph where we omitted the parts that are not relevant for understanding our proof.)
A snake-shaped line corresponds to a path of $0$-weight edges.
The orange vertices and edges are added to $G$ to form graph $H$.
In each block $G_i$, vertex $s_1^{(i)}$ is either connected to $t_1^{(i)}$ or $t_2^{(i)}$.
As shown in \cite{Y-soda21}, the graph induced by these blocks (i.e. without $a$ and $b$) is connected if and only if the latter case happens in an odd number of blocks.
Since all edges except $\set{a,b}$ have weight $0$, the MST of $H$ will contain $\set{a,b}$ if and only if the subgraph induced by $G_1$, $G_2$, and $G_3$ is disconnected.
}
\label{fig:clique_lb_figure}
\end{figure*}
 
Consider any $\clique_1$ algorithm $\mathcal{A}$ that correctly outputs an MST with probability $\tfrac{3}{4}$, after just one round, while sending at most $L$ bits over each link. 
  We now describe how to use $\mathcal{A}$ in the \emph{distributed graph sketching} ($\sketch$) model to solve graph connectivity, where the input graph is assigned to $n$ players, such that each player obtains one vertex and the incident edges as its input.
  The referee, on the other hand, receives a single message (``sketch'') from each player and must compute the output.

  Our lower bound graph $H$ contains a certain subgraph $G$, which corresponds to the construction of Yu~\cite{Y-soda21} and
  was used to show that a sketch length of $\Omega(\log^3 n)$ bits is required for deciding graph connectivity in the $\sketch$ model.
  In more detail, graph $G$ consists of $\beta = \Theta\lb(\sqrt{n}\rb)$ interconnected blocks $G_0,\dots,G_{\beta}$, comprising $\Theta\lb(\sqrt{n}\rb)$ vertices each.
  A block $G_i$ contains four special vertices, $s_1^{(i)}$, $s_2^{(i)}$, $t_1^{(i)}$, and $t_2^{(i)}$ that connect $G_i$ to $G_{i+1 \bmod \beta}$.

  For simulating $\mathcal{A}$, we define $H$ to consist of a copy of $G$ and two additional nodes $a$ and $b$ with fixed unique IDs $id(a)$ and $id(b)$.
  Node $a$ is connected to $s_1^{(1)} \in G_1$, and $b$ is connected to $s_2^{(1)} \in G_1$.
  Moreover, there is an edge between $a$ and $b$.
  We assign weight $0$ to all edges except for the edge $\set{a,b}$, which has weight $1$.
  Figure~\ref{fig:clique_lb_figure} shows the resulting graph, assuming just three blocks.

  Our proof makes use of the following properties of graph $G$:
  \begin{lemma}[\cite{Y-soda21}] \label{lem:ysoda21-graph}
    The graph $G$ has at most $2$ components.
    Vertices $s_1^{(1)}$ and $s_2^{(1)}$ lie in distinct components if and only if $G$ is not connected.
  \end{lemma}
  From the definition of our graph $H$, we are able to tie together the connectivity of $G$ with the inclusion of the edge $\set{a,b}$ in the MST of $H$:

  \begin{observation} \label{obs:equivalence}
    Graph $H$ is connected. Moreover, any MST of $H$ contains the edge $\set{a,b}$ if and only if the subgraph $G$ induced by the vertices in $V(H) \setminus \set{a,b}$  is disconnected.
  \end{observation}

  We now describe the details of simulating the given $\clique_1$ algorithm $\mathcal{A}$ in the $\sketch$ model:
  To determine the messages sent in the first (and only) round, each node simulates the local computation of $\mathcal{A}$ for its assigned vertex of $H$, by taking into account its neighborhood information and the shared randomness.
  Note that the referee knows the neighborhoods of nodes $a$ and $b$,
  because the IDs of their neighbors $s_1^{(1)}$ and $s_2^{(1)}$ are fixed in advance (see Sec.~4.1 in \cite{Y-soda21}).
  As a result, the referee can locally simulate the first (and only) round of algorithm $\mathcal{A}$ for nodes $a$ and $b$. 

  Every other node $u$ simply discards all messages produced by the simulation of $\mathcal{A}$, except for the messages that are addressed to either $a$ or $b$, which $u$ sends as its sketch to the referee instead.
  Recall that $u$ knows the IDs of these vertices since they are fixed in advance.
  Moreover, the messages that are produced by simulating  algorithm $\mathcal{A}$ for a vertex $v$ only depend on $v$'s neighborhood and the (shared) randomness, both of which are readily available to the node in the $\sketch$ model who is responsible for $v$.

  After receiving all messages from the players, the referee has learned all messages that were sent to $a$ and $b$, and it also knows the complete neighborhood information of these vertices.
  Thus, the referee can correctly simulate the output that $a$ and $b$ produce in the $\clique_1$ model after receiving their first round messages.
  Finally, the referee answers ``connected'' if and only if the edge $\set{a,b}$ is not included in the MST edges output by $a$ and $b$.

  It follows from Lemma~\ref{lem:ysoda21-graph}, Observation~\ref{obs:equivalence}, and the fact that $\mathcal{A}$ computes an MST with probability at least $\frac{3}{4}$ that the referee is able to correctly decide whether $G$ is connected with precisely the same probability.
  Theorem~1 in \cite{Y-soda21} tells us that the worst case sketch length of any protocol that solves connectivity in $\sketch$ is $\Omega(\log^3 n)$.
  We conclude that the link bandwidth used by algorithm $\mathcal{A}$ must be at least $\Omega(\log^3 n)$ bits.

\section{Discussion and Open Problems}
\label{sec:future}

Our work suggests several interesting avenues for future exploration.
Since we have shown that multiple rounds are necessary for computing an MST in the congested clique with $O(\log n)$ bandwidth.
On the other hand, the recent line of work \cite{hegeman2015toward,GP-podc16} culminated in the $O(1)$-rounds algorithm of \cite{JN-soda18} for computing an MST in the unicast congested clique.
In fact, \cite{GN-podc18} prove that we can solve even $n^{1-\epsilon}$ many MST instances in $O(1)$ rounds.
These results raise the question regarding the precise dependency between the necessary link bandwidth and the number of rounds.
\begin{openproblem}
  Is there an algorithm that computes an MST in just $2$ rounds of the unicast congested clique with $O(\log n)$ bandwidth?
  Alternatively, is there a tradeoff between the number of rounds and the link bandwidth?
\end{openproblem}

A problem closely related to constructing a spanning tree is testing the connectivity of the graph.
Several previous works have considered this problem in the congested clique, see \cite{pai2020connectivity,jurdzinski2017brief,JN-soda18}.
In the congested clique, graph connectivity has been studied under the assumption that \emph{all} nodes need to output ``yes'' if the graph is connected and at least \emph{one} node outputs ``no'' otherwise.
Under this assumption, the lower bound technique of this paper extends to graph connectivity even for the unicast congested clique.
However, it would be interesting to study connectivity under the more realistic assumption that only \emph{some} node outputs the decision while all other nodes may remain undecided.
Algorithms for this setting can truly leverage the power of the \emph{unicast} congested clique:

\begin{openproblem}
  What is the minimum link bandwidth required for graph connectivity in the one-round unicast congested clique, for algorithms where some arbitrary node outputs the decision?
\end{openproblem}

\bibliographystyle{ACM-Reference-Format}
\bibliography{references}

\end{document}